# FINAL MUON EMITTANCE EXCHANGE IN VACUUM FOR A COLLIDER[*]

Don Summers[†], John Acosta, Lucien Cremaldi, Terry Hart, Sandra Oliveros,
Lalith Perera, Wanwei Wu, University of Mississippi - Oxford, University, MS 38677, USA
David Neuffer, Fermilab, Batavia, IL 60510, USA

## Abstract

We outline a plan for final muon ionization cooling with quadrupole doublets focusing onto short absorbers followed by emittance exchange in vacuum to achieve the small transverse beam sizes needed by a muon collider. A flat muon beam with a series of quadrupole doublet half cells appears to provide the strong focusing required for final cooling. Each quadrupole doublet has a low $\beta$ region occupied by a dense, low Z absorber. After final cooling, normalized xyz emittances of (0.071, 0.141, 2.4) mm-rad are exchanged into (0.025, 0.025, 70) mm-rad. Thin electrostatic septa efficiently slice the bunch into 17 parts. The 17 bunches are interleaved into a 3.7 meter long train with RF deflector cavities. Snap bunch coalescence combines the muon bunch train longitudinally in a 21 GeV ring in 55 $\mu$s, one quarter of a synchrotron oscillation period. A linear long wavelength RF bucket gives each bunch a different energy causing the bunches to drift until they merge into one bunch and can be captured in a short wavelength RF bucket with a 13% muon decay loss and a packing fraction as high as 87%.

## INTRODUCTION

Due to s-channel production, a muon collider [1] may be ideal for the examination of H/A Higgs scalars which could be at the 1.5 TeV/$c^2$ mass scale and are required in supersymmetric models [2]. But what is the status of muon cooling? As noted in Table 1, more than five orders of magnitude of muon cooling have been shown in two simulated designs [3, 4] but not quite the six orders of magnitude needed for a high luminosity muon collider. Also as can be seen in Table 1, some of the longitudinal cooling needs to be exchanged for lower transverse emittance.

A long solenoid [5] with a 14 Tesla magnetic field and a 200 MeV/c muon beam gives a betatron function of $\beta_\perp = 2 p/[3.0 B] = 2(200\,\text{MeV/c})/[3.0\,(14\,\text{T})] = 9.5$ cm and works with high pressure or liquid hydrogen absorber. The short 14 T solenoids in the final stage of the Rectilinear Cooling Channel [4] give $\beta^*$ = 3.1 cm in a region of limited length, which is filled with lithium hydride absorber. To get to the lower betatron values of about 1 cm needed by a muon collider for final cooling, quadrupole doublet cells are explored herein. Table 2 gives transverse equilibrium emittances for a number of low Z materials, particularly those with high densities.

___________
[*] Work supported by NSF Award 0969770
[†] summers@phy.olemiss.edu

Table 1: Helical and Rectilinear Cooling Channel normalized 6D emittances from simulations and the emittance needed for a muon collider. The channels cool by over five orders of magnitude and need less than a factor of 10 more for a collider. The 21 bunches present after initial phase rotation are also merged into one bunch during cooling.

|                      | $\epsilon_x$ (mm) | $\epsilon_y$ (mm) | $\epsilon_z$ (mm) | $\epsilon_{6D}$ (mm$^3$) |
|----------------------|-------|-------|-------|---------|
| Initial Emittance [4]| 48.6  | 48.6  | 17.0  | 40,200  |
| Helical Cooling [3]  | 0.523 | 0.523 | 1.54  | 0.421   |
| Rectlinear Cooling [4]| 0.28 | 0.28  | 1.57  | 0.123   |
| Muon Collider [1]    | 0.025 | 0.025 | 70    | 0.044   |

Table 2: Muon equilibrium emittance at 200 MeV/c ($\beta$ = v/c = 0.88) for hydrogen, lithium hydride, beryllium, boron carbide, diamond, and beryllium oxide absorbers, [6–8]. $\epsilon_\perp = \beta^* E_s^2/(2 g_x \beta m_\mu c^2 (dE/ds) L_R)$, where $\beta^*$ twiss is 1 cm, $E_s$ is 13.6 MeV, the transverse damping partition number $g_x$ is one with parallel absorber faces, $m_\mu c^2$ is 105.7 MeV, and L$_R$ is radiation length.

| Material | Density g/cm$^3$ | L$_R$ cm | dE/ds MeV/cm | $\epsilon_\perp$ (mm - rad) (equilibrium) |
|----------|---------|---------|--------|---------|
| H$_2$ gas | 0.000084 | 750,000 | 0.00037 | 0.036 |
| Li H     | 0.82    | 97      | 1.73   | 0.059 |
| Be       | 1.85    | 35.3    | 3.24   | 0.087 |
| B$_4$C   | 2.52    | 19.9    | 4.57   | 0.109 |
| Diamond  | 3.52    | 12.1    | 6.70   | 0.123 |
| Be O     | 3.01    | 13.7    | 5.51   | 0.132 |

## QUADRUPOLE DOUBLET COOLING

Following Feher and Strait [9] and their paper including the LHC final focus quadrupole triplet design in 1996 with a $\beta^*$ of 50 cm, we look into a short quadrupole doublet for final muon cooling with a $\beta^*$ of 1 cm. Focal lengths in x and y differ in the doublet. The outer two LHC quadrupoles are focusing in the first transverse dimension and defocusing in the second transverse dimension. The inner double length quadrupole is focusing in the second transverse dimension and defocusing in the first transverse dimension. The relation between $\beta$ functions, focal length ($L_f$), and beam size is

$$\beta^* \beta_{\max} = b\,L_f^2 \quad \text{and} \quad \sigma_x = \sqrt{\epsilon_x \beta_x/(\beta\,\gamma)} \qquad (1)$$

where b is a fudge factor equal to 1.65 for the LHC. Thus a lower $\beta^*$ leads to a larger $\beta_{\max}$ and larger bore quadrupoles.

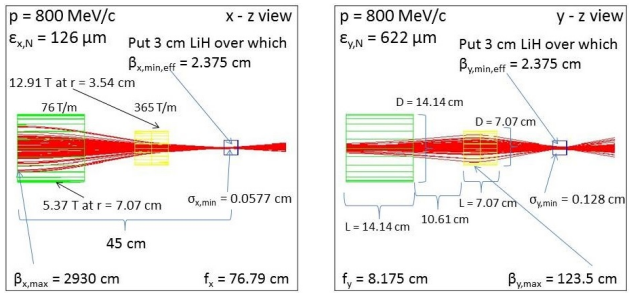

Figure 1: Quadrupole doublet half cell for final muon cooling with a flat beam, $\beta^*_{x,y} = 2$ cm, and a 3 cm long LiH absorber. The G4beamline [10] transmission is 998/1000 and the coverage for quadrupoles is at least $\pm 3.2\sigma$. Coverage closer to $\pm 4\sigma$ would be better.

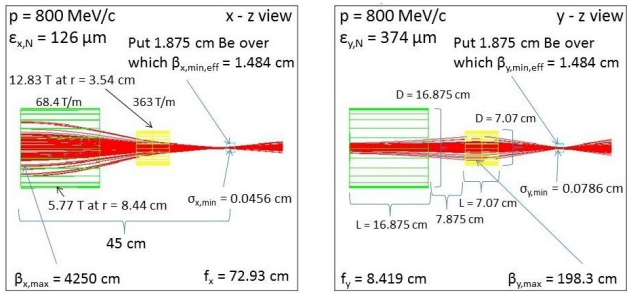

Figure 2: Quadrupole doublet half cell for final muon cooling with a flat beam, $\beta^*_{x,y} = 1.25$ cm, and a 1.875 cm long berylium absorber. The G4beamline transmission is 998/1000 and the coverage for quadrupoles is at least $\pm 3.2\sigma$.

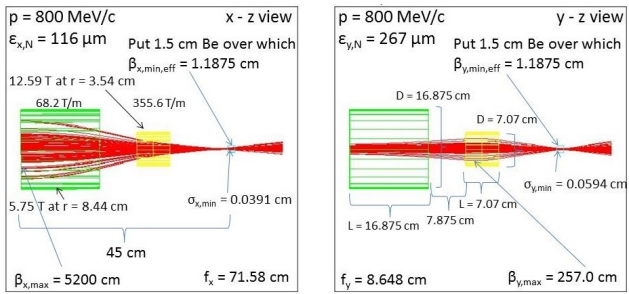

Figure 3: Quadrupole doublet half cell for final muon cooling with a flat beam, $\beta^*_{x,y} = 1$ cm, and a 1.5 cm long berylium absorber. The G4beamline transmission is 996/1000 and the coverage for quadrupoles is at least $\pm 3.0\sigma$.

A short length of low Z absorber absorber is placed at the focus of each quadrupole doublet as shown in Fig. 1 to 5. Flat beams are used with the $\cos(2\theta)$ quadrupole doublets which do not exceed 14 T as in the LHC Nb$_3$Sn LARP quadrupoles [11]. Flat beams might be generated either by slicing with a septum followed by recombination [12] or, if angular momentum can be added to the beam, by a skew quadrupole triplet [13]. The round to flat triplet transformation has been successfully simulated for muons [14]. Note that $\beta(s) = \beta^* + s^2/\beta^*$. As $\beta^*$ becomes smaller, the absorber must become thinner in the beam direction $s$. An OREO cookie geometry appears to be useful with a denser absorber on the inside and a lower Z absorber on the outside. The fringe fields [15] of the magnet fall off as the distance cubed which may help to ameliorate RF breakdown. The beam power of $4 \times 10^{12}$ 800 MeV/c muons (KE = 701 MeV) arriving at 15 Hz is 6700 watts of which only a tiny fraction would heat the superconductor.

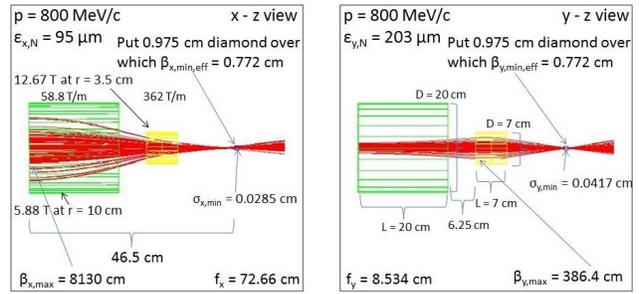

Figure 4: Quadrupole doublet half cell for final muon cooling with a flat beam, $\beta^*_{x,y} = 0.65$ cm, and a 0.975 cm long diamond absorber. The G4beamline transmission is 998/1000 and the coverage for quadrupoles is at least $\pm 3.1\sigma$.

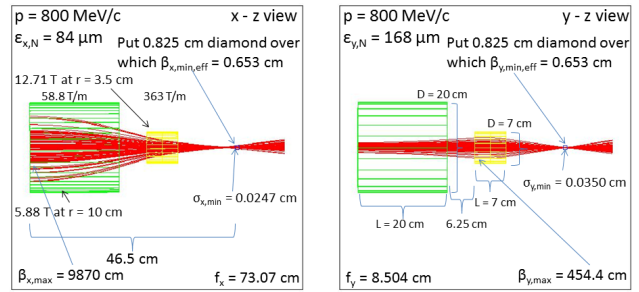

Figure 5: Quadrupole doublet half cell for final muon cooling with a flat beam, $\beta^*_{x,y} = 0.55$ cm, and a 0.825 cm long diamond absorber. The G4beamline transmission is 995/1000 and the coverage for quadrupoles is at least $\pm 3.0\sigma$.

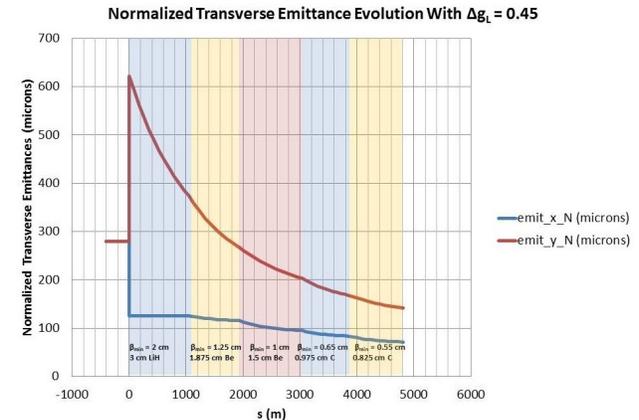

Figure 6: Transverse emittance versus channel length. The 6D xyz muon beam emittance is first transformed from (0.280, 0.280, 1.57) to (0.126, 0.622, 1.57) mm-rad and then cooled in five stages to (0.0714, 0.141, 2.418) mm-rad to meet the requirements of a muon collider. The total five stage channel length is 4808 meters and 62% of the 800 MeV/c ($\gamma = 7.63$) muons decay. Less decay would be better.

For each stage, cooling [6] is calculated (see Table 3 and Fig. 6) from the equilibrium emittances, the characteristic transverse and longitudinal cooling lengths, and the cooling partition numbers, $g_{xyz}$ = (1.0, 0.55, 0.57). The cooling lengths are inversely proportional to their corresponding partition numbers. The longitudinal equilibrium emittance depends on the longitudinal betatron function, energy, longitudinal partition number, and absorber ionization energy. The longitudinal betatron function depends on the RF wavelength, energy, RF gradient averaged over the cooling cell length, and the RF phase.

Table 3: The five cooling stages. $\phi_0$ is the angle away from the 650 MHz, 25 MV/m, 2.5 meter long RF zero crossing. $\phi_0$ is set so that the RF replaces the energy lost in the absorber. The RF is long to limit longitudinal emittance increase. The empty space between quadrupoles and RF is 17.5 cm. The damping partition numbers $g_{xyx}$ are (1.0, 0.55, 0.57). The absorber wedge angle assumes a dispersion, $\eta$ = 1 cm. The carbon is diamond. The emittances are normalized.

| Stage | 1 | 2 | 3 | 4 | 5 |
|---|---|---|---|---|---|
| No. of Cells | 281 | 235 | 294 | 215 | 254 |
| Cell Length (m) | 3.75 | 3.75 | 3.75 | 3.78 | 3.78 |
| Stage Length (m) | 1052 | 882 | 1102 | 812 | 960 |
| Doublet Length (cm) | 45 | 45 | 45 | 46.5 | 46.5 |
| RF Phase $\phi_0$ | $4.72^0$ | $5.63^0$ | $4.50^0$ | $6.16^0$ | $5.21^0$ |
| Wedge Angle | $68.0^0$ | $45.8^0$ | $37.3^0$ | $24.7^0$ | $21.0^0$ |
| $\beta^*_{xy}$ (cm) | 2.0 | 1.25 | 1.0 | 0.65 | 0.55 |
| Absorber | LiH | Be | Be | C | C |
| Absorber Length (cm) | 3.0 | 1.875 | 1.5 | 0.975 | 0.825 |
| Quad Doublet Bore $\sigma$ | ±3.2 | ±3.2 | ±3.0 | ±3.1 | ±3.0 |
| Quad Doub. Gap (cm) | 10.61 | 7.875 | 7.875 | 6.25 | 6.25 |
| Final $\epsilon_x$ (mm-rad) | 0.126 | 0.116 | 0.095 | 0.084 | 0.071 |
| Final $\epsilon_y$ (mm-rad) | 0.374 | 0.268 | 0.203 | 0.168 | 0.141 |
| Final $\epsilon_z$ (mm-rad) | 2.058 | 2.273 | 2.347 | 2.400 | 2.418 |

In summary, quadrupole doublets and dense, low Z absorbers cool the current outputs given either by the Helical [3] or Rectilinear [4] 6D muon cooling channels and prepare the input for emittance exchange which reduces the normalized transverse beam emittance to the 0.025 mm emittance required by a high luminosity muon collider. A linear channel is shown here, but the emittance is low enough to inject into rings, where the momentum compaction might lower longitudinal emittance.

## SLICE THE BEAM WITH SEPTA

Slice [16] the muon beam into 17 pieces with two septa vertically and then 14 septa horizontally. A $w$ = 0.1 mm wide electrostatic septa gives a total fractional loss of

$$\frac{4\sqrt{2}\,w}{x_{\max}} = \frac{4\sqrt{2} \times 0.1\,\text{mm}}{17\,\text{mm}} = 0.03, \quad (2)$$

where $x_{\max}$ is a 300 MeV/c beam size (equation 3).

$$\sqrt{\frac{\epsilon_{N,x}\,\beta_x}{\beta\,\gamma}} = \sqrt{\frac{(0.100\,\text{mm})(8,000\,\text{mm})}{2.84}} = 17\,\text{mm} \quad (3)$$

## CREATE A 3.7 M LONG BUNCH TRAIN WITH RF DEFLECTOR CAVITIES

Combine 17 bunches into a 3.7 m long train with RF deflector cavities as used in CLIC tests. Each cavity interleaves two or three bunch trains. Deflection is ±4.5 mrad or zero at 300 MeV/c [17]. The final train has a 231 mm bunch spacing for acceleration by 1300 MHz RF cavities (see Table 4). Estimate the required kick [18] to inject a 300 MeV/c ($\gamma\beta$ = 300/105.7 = 2.84) beam with a normalized emittance of 0.025 mm-rad and a $\beta_\perp$ of 8000 mm. The kick must be 4x greater than the rms divergence of the beam or $4\sqrt{\epsilon/(\gamma\beta\beta_\perp)}$ = 4.2 mrad, which matches CLIC. The $\pm 4\sigma$ beam diameter is $8\sqrt{\epsilon\beta_\perp/(\gamma\beta)}$ = 67 mm.

## SNAP BUNCH COALESCE A TRAIN OF 17 BUNCHES INTO ONE IN A RING

Finally, *snap bunch coalescing* with RF is used to combine the 17 muon bunches longitudinally. In snap bunch coalescing, all bunches are partially rotated over a quarter of a synchrotron period in energy-time space with a linear long wavelength RF bucket and then the bunches drift in a ring until they merge into one bunch and can be captured in a short wavelength RF bucket. The bunches drift together because they each have a different energy set to cause the drift. Snap bunch coalescing replaced adiabatic bunch coalescing at the Fermilab Tevatron collider program and was used for many years [19]. Sets of fifteen bunches were combined in the Tevatron. A 21 GeV ring has been used in a simulation [20] with ESME [21] to show the coalescing of 17 muon bunches in 55 $\mu$s. The muon decay loss was 13%. The lattice had $\gamma_t$ = 5.6 [22]. The RF frequencies were 38.25 MHz and 1.3 GHz and the longitudinal packing fraction was as high as 87% [23]. The initial normalized 2.4 mm longitudinal emittance is increased by a factor of 17/0.87 to become 47 mm, which is less than the 70 mm needed for a muon collider and allows for some dilution.

Table 4: Combine 17 bunches into a 3.7 m long train with 10 RF Deflector Cavities. Each cavity interleaves two or three bunch trains. Deflection is ±4.5 mrad or zero at 300 MeV/c. The RF deflection frequencies used are 731, 487, and 650 MHz. The final train has a 231 mm bunch spacing for acceleration by 1300 MHz RF cavities.

| Number of Trains Interleaving | Number of RF Cavities | RF Wavelength | Output Spacing in Wavelengths | Output Bunch Spacing |
|---|---|---|---|---|
| 17 → 6 | 6 | 410 mm | 9/4 | 923 mm |
| 6 → 2 | 3 | 616 mm | 3/4 | 462 mm |
| 2 → 1 | 1 | 462 mm | 1/2 | 231 mm |